\documentclass[final]{siamart0516}
\usepackage{comment}

\usepackage{lipsum}
\usepackage{amsfonts}
\usepackage{amsmath}
\usepackage{amssymb}
\usepackage{graphicx}
\usepackage{epstopdf}
\usepackage{algorithmic}
\usepackage{physics}
\usepackage{optidef}
\usepackage{bbm}
\ifpdf
  \DeclareGraphicsExtensions{.eps,.pdf,.png,.jpg}
\else
  \DeclareGraphicsExtensions{.eps}
\fi

\usepackage{amsmath,amssymb} 
\usepackage{enumerate}

{\bf}{} 

\newcommand{\TheTitle}{Spectral Methods for Quantum Optimal Control: Artificial Boundary Conditions} 
\newcommand{\TheAuthors}{
A. Wodecki, G. Korpas, J. Marecek}

\headers{Spectral Methods for Quantum Optimal Control}{\TheAuthors}

\title{{\TheTitle}}

\author{
  Ales Wodecki\thanks{Department of Computer Science, Czech Technical University in Prague, Czech Republic. Contact: \email{wodecki.ales@fel.cvut.cz}.} \and Jakub Marecek\thanks{Department of Computer Science, Czech Technical University in Prague, Czech Republic.} \and Vyacheslav Kungurtsev\thanks{Department of Computer Science, Czech Technical University in Prague, Czech Republic.} \and Pavel Eichler\thanks{Faculty of Nuclear Sciences and Physical Engineering, Czech Technical University in Prague, Czech Republic} \and Georgios Korpas\thanks{Quantum Technologies Group, Innovation \& Ventures, HSBC Lab, London, UK.} \and Philip Intallura\thanks{Quantum Technologies Group, Innovation \& Ventures, HSBC Lab, London, UK.}
}

\usepackage{amsopn}

\usepackage{mathtools}

\usepackage{comment}
\usepackage{amsmath}

\ifpdf
\hypersetup{
  pdftitle={\TheTitle},
  pdfauthor={\TheAuthors}
}
\fi

\begin{document}

\maketitle

\begin{abstract}
  The problem of quantum state preparation is one of the main challenges in achieving the quantum advantage. Furthermore, classically, for multi-level problems, our ability to solve the corresponding quantum optimal control problems is rather limited. The ability of the latter to feed into the former may result in significant progress in quantum computing. To address this challenge, we propose a formulation of quantum optimal control that makes use of artificial boundary conditions for the Schr\"odinger equation in combination with spectral methods. The resulting formulations are well suited for investigating periodic potentials and lend themselves to direct numerical treatment using conventional methods for bounded domains.
\end{abstract}


\section{Introduction}

Quantum optimal control \cite{d2021introduction} has a number of important applications, including initial state preparation and gate implementation quantum computing \cite{peirce1988optimal,Krotov1993,glaser2015training,DAlessandro2021-dp,magann2021pulses}, or laser control of chemical reactions \cite{135209,Assion1998,zare1998laser,PhysRevLett.68.1500,werschnik2007quantum,balint2008optimal}. All of this is of considerable and growing importance.

In the context of quantum computing, certain quantum algorithms promise substantial speedup, over classical counterparts, in a variety of applications including quantum alternatives to Monte Carlo simulations \cite{intallura2023survey} or quantum optimization \cite{abbas2023quantum}. However, to preserve these speedups, the initial quantum state preparation (or distribution) \cite{kak1999initialization} must be carried out in a unit amount of time. When the complexity of initial quantum state preparation using one controlled rotation per scalar, also known as the Grover-Rudolph state preparation \cite{grover2002creating}, is considered, the speed-up may vanish. A prime example includes quantum algorithms replacing classical Monte Carlo simulations \cite{intallura2023survey}, where Grover-Rudolph state preparation is known to remove the quadratic speed-up even in the case of log-concave probability distributions \cite{herbert2021no}. In general, loading an arbitrary quantum state over $n$-qubit registers requires $O(4^n)$ CNOT gates \cite{Grover2000, plesch2011quantum} which conflate any potential for useful speed-up.
It is hence natural to seek alternative methods for initial quantum-state preparation. Although several interesting approaches have been suggested in the literature, the most widely used generative approaches \cite{Zoufal2019,PhysRevResearch.4.043092} do not necessarily scale as desired, while parametrized circuit approaches often suffer from the typical problems associated with them, that is barren plateaus \cite{wang2021noise,larocca2022diagnosing,schumann2023emergence, ragone2023unified, diaz2023showcasing} and while several approaches have been proposed to (partially) overcome this problem \cite{Friedrich_2022,robertson2022escaping,zhang2022escaping,mastropietro2023fleming} these methods, to the best of our knowledge, have not been tested in the context of the state preparation problem. 

Diving deeper, within the quantum circuit model in quantum computing, the performance depends crucially on the fidelity of two-qubit gates \cite{Zajac2018,Wright2019}. These are typically implemented using microwave or laser pulses, which are computed using quantum optimal control \cite{Koch2022}. While it has been traditionally suggested that performance of local-search methods such as GRAPE \cite{Khaneja2005} and CRAB \cite{Caneva_2011} applied to simple two-qubit Hamiltonians is sufficient, it is increasingly recognized that one could improve the performance substantially by considering methods \cite[e.g.]{bondar2022quantum} that provably converge to the global optima and, perhaps even more importantly, one should like to consider open quantum systems, rather than closed quantum systems. In the latter, to address issues such as cross-talk, one may need to consider a multi-qubit system, in order to design the pulse implementing a two-qubit model. 

Outside of quantum computing, in applications such as the laser control of chemical reactions \cite{zare1998laser,balint2008optimal} or coherent population transfer among quantum states of atoms and molecules \cite{vitanov2001laser,bergmann1998coherent} more broadly, one would naturally like to implement quantum optimal control of many-level systems, but even popular heuristics such as GRAPE and CRAB do not scale sufficiently. 

In this article, we address the scalability of quantum optimal control methods by considering a spectral-domain (or frequency-domain) approach \cite{friesecke2018frequency,aerts2022laser}. 
In their pioneering work, 
Friesecke \emph{et al}.\  \cite{friesecke2018frequency} provide structural results pertaining to the time-frequency transformation of a quantum system, providing time-frequency formulated constraints along with an alternative cost function regularization, which penalizes with respect to the frequency space. Their results provide structural underpinnings for time-frequency methods in general and may be directly applied to systems with a finite-level Hamiltonian. Further validating this approach are examples of physical systems which have a finite number of components in frequency space. Some examples of such systems are spin-1/2 particles in a magnetic field or electronic states of atoms in laser fields. Numerically, \cite{friesecke2018frequency}  have shown that this setting leads to frequency-space sparse controls in many cases. 
Building on the aforementioned approach, we develop a time-frequency formulation which allows for direct numerical treatment even for infinite-level Hamiltonians. This further develops the ideas in \cite{friesecke2018frequency} making them applicable to a wider class of problems.

In particular, we propose a formulation of the control problem that makes use of \emph{artificial boundary conditions} \cite{BASKAKOV1991123,Dalrymple1992PerfectBC,sofronov_1998} for the Schrödinger equation in combination with the aforementioned time-spectral methods. The resulting formulations are well suited for investigating periodic potentials and lend themselves to direct numerical treatment using conventional methods for bounded domains. This could be seen as a harmonic-analytic approach to quantum optimal control, inasmuch as we utilize the Laplace transform and the inverse Laplace transform to control the complexity of the problem solved classically in order to control the quantum system optimally. 

\section{Problem Formulation}

There are many numerical treatments that lead to the successful solution of the Schrödinger equation \cite{Abdurrouf_2021, Salehi20, Becke90}. Several of these methods are centered on finding ways to reduce the infinite spatial domain, which is inherent in the problem \cite{Orszag77, bigReviewABC, Ladouceur96}. In particular, pseudospectral methods make use of a Fourier basis of test functions to discretize both the temporal and spatial domains, resulting in a fully discrete problem \cite{Borz2017FormulationAN}. Another noteworthy approach, called the artificial boundary method, is inspired by classical numerics \cite{sofronov_1998, bigReviewABC, Frensley90}. In this method, one attempts to impose artificial boundary conditions on the Schrödinger equation, making it solvable using conventional numerical methods (finite elements, finite differences, etc.). Imposing these artificial boundary conditions finds utility in a broad range of quantum and non-quantum applications \cite{papa11, Dalrymple1992PerfectBC, BASKAKOV1991123}, including non-Markovian open quantum systems \cite{Hellums94}. For completeness, we note that, in rare cases, it is possible to transform the original problem domain into a bounded one and avoid the construction of these artificial boundary conditions \cite{Orszag77, Ladouceur96}. In the following, a hybid method is presented, combining both of these ``schools of thought''.

Although the presented techniques apply to the spatial domain of any fixed dimenstion \cite{Yao18, bigReviewABC}, i.e. to $\mathbb{R}^{n}$, the presentation is kept simple by assuming that $n=1$. Let $\psi_{\text{ini}}\in L^{2}\left(\mathbb{R}\right)$ be a compactly supported initial condition, and let $V:\mathbb{R}\times\mathbb{R}_{0}^{+}\rightarrow\mathbb{R}$ be a potential that satisfies 

\begin{equation}
    V\left(x\right)=\begin{cases}
V_{l} & \text{ for }x\leq x_{l}\\
V_{r} & \text{ for }x\geq x_{r},
\end{cases}
\end{equation}
where $x_{l}, x_{r} \in \mathbb{R}$. The quantum system of interest is described by the Schrödinger equation

\begin{equation}
  \begin{aligned}
    i\frac{\partial\psi\left(x,t\right)}{\partial t}=-\frac{\partial^{2}\psi\left(x,t\right)}{\partial x^{2}}+V\left(x,t\right)\psi\left(x,t\right), && \text{where }\left(x,t\right)\in\mathbb{R}\times\mathbb{R}_{0}^{+}, \\
    \underset{\left|x\right|\rightarrow+\infty}{\lim}\psi\left(x,t\right)=0, && \text{ for all }t\in\mathbb{R}_{0}^{+}, \\
    \psi\left(x,0\right)=\psi_{\text{ini}}\left(x\right), && \text{ for all }x\in\mathbb{R}.\label{Schrodinger_auto}
  \end{aligned}
\end{equation}
The control is introduced in Eq. \eqref{Schrodinger_auto} by assuming that $V\left(x,t\right)=V\left(x,t;\eta\right),$ where $\eta$ is a finite or infinite-dimensional control parameter that belongs to the set $C$. Additionally, assume that there exists $\widehat{x}_{l}, \widehat{x}_{r}\in \mathbb{R}$ such that 

\begin{equation}
\text{for all }\eta\in C\text{ , }V\left(x,t;\eta\right)\text{ is constant on }\left(-\infty,\widehat{x}_{l}\right)\text{ and \ensuremath{\left(\widehat{x}_{r},+\infty\right)}}.\label{key_condition_to_pinpoint_BC}
\end{equation}
The above condition allows us to fix the spatial coordinates at which the artifical boundary conditions are imposed. 

\subsection{The Possible Formulations of the Control Set}
Using the setting described in the previous section, one can define two very distinct ways in which the control can be imposed. To derive the first one of these, let us assume that there is no a-priori assumption on the potential $V$, while still imposing (\ref{key_condition_to_pinpoint_BC}). Using the standard existence result for the Schrödinger equation (add a link to the appendix and ref), we may consider the following control sets 
\begin{equation}
\begin{aligned}
C_{\text{hom}}^{\widehat{x}_{l},\widehat{x}_{r}}=\left\{ V\in C\left(\mathbb{R}_{0}^{+},L^{\infty}\left(\mathbb{R}\right)\right):V\left(x,t\right)=0\text{ for }x\in\left(-\infty,\widehat{x}_{l}\right)\cup\left(\widehat{x}_{r},+\infty\right)\forall t\in\mathbb{R}_{0}^{+}\right\} , \\
C^{\widehat{x}_{l},\widehat{x}_{r}}=\left\{ V\in C\left(\mathbb{R}_{0}^{+},L^{\infty}\left(\mathbb{R}\right)\right):V\text{ is constant on }\left(-\infty,\widehat{x}_{l}\right)\text{ and }\left(\widehat{x}_{r},+\infty\right)\forall t\in\mathbb{R}_{0}^{+}\right\}, \label{control_1}
\end{aligned}
\end{equation}
where notably both of these sets are convex subsets of the Banach space $C\left(\mathbb{R}_{0}^{+},L^{\infty}\left(\mathbb{R}\right)\right)$, which ensures the existence of a solution in $C\left(\mathbb{R}_{0}^{+},L^{2}\left(\mathbb{R}\right)\right)$ for any such control \cite{hinze09}.

In the second approach, we assume that the potential has a given structure based on some knowledge of the system. A simple, but illustrative example of this is a driven quantum harmonic oscillator, which has the potential
\begin{equation}
V\left(x,t\right)=\frac{\omega^{2}\left(t\right)q^{2}\left(x,t\right)}{2m}-\widetilde{J}\left(x,t\right)q\left(x,t\right),\label{driven_harmonic_oscilator_potential}
\end{equation}
where $\omega$ denotes the angular velocity, $q$ is the position operator and $\widetilde{J}$ is a bounded function, which describes the corrected drive of the oscilator. The correction is incorporated so that condition (\ref{key_condition_to_pinpoint_BC}) is satisfied, i.e. 
\begin{equation}
    \widetilde{J}\left(x,t\right)=J\left(t\right)+R\left(x,t\right),
\end{equation}
where $J$ is the physically acurate drive and 
\begin{equation}
    R\left(x,t\right)=\begin{cases}
\frac{1}{x\left(t\right)}\left[x_{r}+\frac{\omega^{2}\left(t\right)x^{2}\left(t\right)}{2m}J\left(t\right)\right], & \text{for }x\in\left(x_{r},+\infty\right)\\
0, & \text{for }x\in\left(x_{l},x_{r}\right)\\
\frac{1}{x\left(t\right)}\left[x_{l}+\frac{\omega^{2}\left(t\right)x^{2}\left(t\right)}{2m}J\left(t\right)\right], & \text{for }x\in\left(-\infty,x_{l}\right)
\end{cases}\label{correction_term}
\end{equation}
is the correction term. Since (\ref{correction_term}) needs to be imposed to ensure that the left and right artificial boundary conditions remain at the spatial points $x_{l}, x_{r}$, respectively, it is clear that the drive $J$ can not have an arbitrary form. To guarantee the existence of soltuion, $J$ must be chosen so that (\ref{driven_harmonic_oscilator_potential}) is of class $C\left(\mathbb{R}_{0}^{+},L^{\infty}\left(\mathbb{R}\right)\right)$. Using this example, we may define the control set as 
\begin{equation}\label{control_2}
    C_{\text{spec}}=\left\{J\in L^{2}\left(\mathbb{R}_{0}^{+}\right):\exists R\in L^{\infty}\left(\mathbb{R}_{0}^{+},L^{\infty}\left(\mathbb{R}\right)\right)\text{ s.t. }V=J+R\in C\left(\mathbb{R}_{0}^{+},L^{\infty}\left(\mathbb{R}\right)\right)\right\},
\end{equation}
where the convexity of such a control set is problem dependant.
\subsection{Possible Extensions of the Formulation}\label{sec_extensions}
The previous section describes how one can impose a control given by a potential in the Schrödinger equation. It is important to note that this does not cover all the possibilities of control. In particular, incorporating the effects of a controllable magnetic field on a quantum harmonic oscilator leads to a change of the momentum operator, i.e.
\begin{equation}
    \frac{\partial^{2}}{\partial x^{2}}\rightarrow\frac{\partial^{2}}{\partial x^{2}}+\eta A,
\end{equation}
where $A$ is the vector potential of the magentic field and $\eta$ is the controlable strength of the magnetic field. Even though this case is not treated in the present article, the generalization is streightforward.

\subsection{Formulation of the Problem Using Artificial Boundary Conditions}
To derive artifical boundary conditions for the problem (\ref{Schrodinger_auto}) the spatial domain is divided into two parts
\begin{equation}
\Omega_{\text{in}}=\left(x_{l},x_{r}\right)\text{ and }\Omega_{\text{out}}=\left(-\infty,x_{l}\right)\cup\left(x_{r},+\infty\right),
\end{equation}
resulting in the coupled system

\begin{equation}
    \begin{aligned}
i\frac{\partial v}{\partial t}+\frac{\partial^{2}v}{\partial x^{2}}=Vv\text{,} && \text{on }\Omega_{\text{in}}\times\mathbb{R}_{0}^{+} \\
\frac{\partial v}{\partial x}=\frac{\partial w}{\partial x}\text{,} && \text{on }\left\{ x_{l},x_{r}\right\} \times\mathbb{R}_{0}^{+} \\
v=\psi_{\text{ini}}\text{,} && \text{on }\Omega_{\text{in}}\times\left\{ 0\right\} \\
i\frac{\partial w}{\partial t}+\frac{\partial^{2}w}{\partial x^{2}}=Vw\text{,} && \text{on }\Omega_{\text{out}}\times\mathbb{R}_{0}^{+} \\
v=w\text{,} && \text{on }\left\{ x_{l},x_{r}\right\} \times\mathbb{R}_{0}^{+}, \\
\underset{\left|x\right|\rightarrow+\infty}{\lim}w\left(x,t\right)=0\text{,} && \text{for all }t\in\mathbb{R}^{+} \\
w=0\text{,} && \text{on } \Omega_{\text{out}}\times\left\{ 0\right\}  
    \end{aligned},\label{split_schrodinger_system}
\end{equation}
where $v$ and $w$ are the respective solutions on the interior $\Omega_{\text{in}}$ and exterior $\Omega_{\text{out}}$ of the full spatial domain $\mathbb{R}$. If $v$ and $\frac{\partial v}{\partial x}$ (or equivalently $w$ and $\frac{\partial w}{\partial x}$) are such that (\ref{split_schrodinger_system}) is well posed we call the boundary conditions absorbing (ABC). If the boundary conditions yield a solution the conincides with the original problem (\ref{Schrodinger_auto}), these boundary conditions are called transparent (TBC).

A transparent boundary condition for problem (\ref{split_schrodinger_system}) can be derived as follows. Let 
\begin{equation}
    \Omega_{\text{l}}=\left(-\infty,x_{l}\right), \quad\Omega_{\text{r}}=\left(x_{r},+\infty\right).
\end{equation}
Transforming the eqution on $\Omega_{\text{r}}$ (given by Eq. (\ref{split_schrodinger_system})) with respect to time using the unilateral Laplace transform we arrive at
\begin{equation}
    is\widehat{w}+\frac{\partial^{2}\widehat{w}}{\partial x^{2}}=V_{r}\widehat{w}\text{, where }x\in\Omega_{\text{r}},s\in\mathbb{C},\label{laplace_of_main_eq}
\end{equation}
where $V_{r}$ denotes the constant value of the potential on $\Omega_{\text{r}}$. The general solution of this ODE reads
\begin{equation}
\widehat{w}\left(x,s\right)=c\left(s\right)^{+}e^{\sqrt{-is+V_{r}}x}+c\left(s\right)^{-}e^{-\sqrt{-is+V_{r}}x}\text{, where }x\in\Omega_{\text{r}},s\in\mathbb{C}.\end{equation}
A physical solution of the equation must be such that the real part of $\sqrt{-is+V_{r}}x$ is postive and so \begin{equation}
    \widehat{w}\left(\cdot,s\right)\in L^{2}\left(\Omega_{\text{r}}\right)\Rightarrow c\left(s\right)^{+}=0.
\end{equation}
The functions $v$ and $w$ are by Eq. (\ref{split_schrodinger_system}) identical on the boundary, and so (using the continuity of the solution up to the boundary) we get
\begin{equation}
    \widehat{w}\left(x,s\right)=e^{\sqrt{-is+V_{r}}\left(x-x_{r}\right)}\widehat{v}\left(x_{r},s\right).\label{boundary_laplace}
\end{equation}
Taking the derivative of (\ref{boundary_laplace}) leads to
\begin{equation}
    \frac{\partial}{\partial x}\widehat{w}\left(x,s\right)\left|_{x=x_{r}}\right.=-\sqrt{-is+V_{r}}\widehat{v}\left(x_{r},s\right)=-\sqrt{-is+V_{r}}\widehat{w}\left(x_{r},s\right).
\end{equation}
An analogous procedure leads to the condition for the left boundary, which reads
\begin{equation}
    -\frac{\partial}{\partial x}\widehat{w}\left(x,s\right)\left|_{x=x_{l}}\right.=-\sqrt{-is+V_{l}}\widehat{v}\left(x,s\right)\left|_{x=x_{l}}\right.=-\sqrt{-is+V_{l}}\widehat{w}\left(x,s\right)\left|_{x=x_{l}}\right.
\end{equation}
Applying the inverse Laplace transform and recalling the inner problem of (\ref{split_schrodinger_system}) yields the formulation
\begin{equation}
    \begin{aligned}
        i\frac{\partial v}{\partial t}+\frac{\partial^{2}v}{\partial x^{2}}=Vv\text{,} && \text{on }\Omega_{\text{in}}\times\mathbb{R}_{0}^{+} \\
\frac{\partial v}{\partial x}=\mathcal{L}^{-1}\left(f\right)\text{,} && \text{on }\left\{ x_{l},x_{r}\right\} \times\mathbb{R}_{0}^{+} \\
v=\psi_{\text{ini}}\text{,} && \text{on }\Omega_{\text{in}}\times\left\{ 0\right\},
    \end{aligned}
\end{equation}
where 
\begin{equation}
    f\left(x,s\right)=\begin{cases}
-\sqrt{-is+V_{r}}\widehat{v}\left(x,s\right)\left|_{x=x_{r}}\right. & x=x_{r}\\
\sqrt{-is+V_{l}}\widehat{v}\left(x,s\right)\left|_{x=x_{l}}\right. & x=x_{l}\label{f_def}
\end{cases}.
\end{equation}
Note that the presented approach is not limited to the case of a constant potential $V$. In particular, the method above has been generalized to linear, piecewise constant and even periodic potentials \cite{Zheng08, PAPADAKIS08, Levy01, Ehrhard08}.

\section{Direct Formulation of the Optimization Problem}
Let $\overline{\psi}\in L^{2}\left(\Omega_{\text{in}}\right)$ denote the target state to be achieved in time $T>0$ and $p, q\in \mathbb{N}$. Using the state equation (\ref{split_schrodinger_system}) with the dependance on the control described by (\ref{control_1}) or (\ref{control_2}) the control problem of interest can be formulated as follows (using (\ref{boundary_laplace}))
\begin{equation}
    \begin{aligned}
        \underset{\eta}{\min}\frac{\alpha}{p}\underset{\Omega_{\text{in}}}{\int}\left|v\left(x,T;\eta\right)-\overline{\psi}\left(x\right)\right|^{p}dx+\beta\frac{1}{q}\underset{\mathbb{R}\times\mathbb{R}_{0}^{+}}{\int}\mu\left(\eta\left(x,t\right)\right)^{q}d\left(x,t\right) \\
        \text{s.t. }i\frac{\partial v}{\partial t}+\frac{\partial^{2}v}{\partial x^{2}}=V\left(\eta\right)v\text{,} && \text{on }\Omega_{\text{in}}\times\mathbb{R}_{0}^{+} \\
\frac{\partial v}{\partial x}=\mathcal{L}^{-1}\left(f\left(\eta\right)\right)\text{,} && \text{on }\left\{ x_{l},x_{r}\right\} \times\mathbb{R}_{0}^{+} \\
v=\psi_{\text{ini}}\text{,} && \text{on }\Omega_{\text{in}}\times\left\{ 0\right\} \\
i\frac{\partial w}{\partial t}+\frac{\partial^{2}w}{\partial x^{2}}=V\left(\eta\right)w\text{,} && \text{on }\Omega_{\text{out}}\times\mathbb{R}_{0}^{+} \\
w=\mathcal{L}^{-1}\left(g\left(\eta\right)\right)\text{,} && \text{on }\left\{ x_{l},x_{r}\right\} \times\mathbb{R}_{0}^{+}, \\
\underset{\left|x\right|\rightarrow+\infty}{\lim}w\left(x,t\right)=0\text{,} && \text{for all }t\in\mathbb{R}^{+} \\
w=0\text{,} && \text{on } \Omega_{\text{out}}\times\left\{ 0\right\}  \label{the_main_optimization_problem}
    \end{aligned}
\end{equation}
where $\mu:C\rightarrow\mathbb{R}_{0}^{+}$ denotes the measure applied to quantify the cost of the control, $\alpha, \beta>0$ denote regularization strengths and 
\begin{equation}
    g\left(\eta\right)=e^{-\sqrt{-is+V_{r}}\left(x-x_{r}\right)}\widehat{v}\left(x_{r},s\right).\label{g_def}
\end{equation}
Assuming the controllability of the system, the first summand of the cost function may be removed and the constraint
\begin{equation}
    \overline{\psi}\left(\cdot,T\right)=v\left(\cdot,T\right)
\end{equation}
imposed. 

It should be noted that the formulation (\ref{the_main_optimization_problem}) is designed to be numerically solvable. For example, one could use a direct optimization method and solve the problem in $\Omega_{\text{in}}$ for a given control $\eta$ and then draw upon the known analytical solution for the problem on $\Omega_{\text{out}}$ (this is possible due to the convenient form of the potential on $\Omega_{\text{in}}$). As we will see in the following section, it remains numerically treatable even after transformation into frequency space.

\section{Spectral Formulation of the Optimization Problem}
The problem (\ref{the_main_optimization_problem}) has been constructed with two objectives in mind:
\begin{itemize}
    \item The direct numerical solvability.
    \item The possibility of reformulation using spectral methods.
\end{itemize}
Using (\ref{f_def}) and (\ref{g_def}) along with (\ref{laplace_of_main_eq}) and analogous computations performed for $\Omega_{\text{in}}$ and $\Omega_{\text{l}}$ lead to the semi-spectral variant of the optimization problem, which reads
\begin{equation}
    \begin{aligned}
        \underset{\eta}{\min}\frac{\alpha}{p}\underset{\Omega_{\text{in}}}{\int}\left|\mathcal{L}^{-1}\left(\widehat{v}\left(s,T;\eta\right)\right)\left(x\right)-\overline{\psi}\left(x\right)\right|^{p}dx+\frac{\beta}{q}\underset{\mathbb{R}\times\mathbb{R}_{0}^{+}}{\int}\mu\left(\eta\left(x,t\right)\right)^{q}d\left(x,t\right) \\
        \text{s.t. }is\widehat{v}+\frac{\partial^{2}\widehat{v}}{\partial x^{2}}=\mathcal{L}\left(V\left(\eta\right)v\right)\text{,} && \text{on }\Omega_{\text{in}}\times\mathbb{R}_{0}^{+} \\ 
        \frac{\partial\hat{v}} {\partial x}=f\left(\eta\right)\text{,} && \text{on }\left\{ x_{l},x_{r}\right\} \times\mathbb{R}_{0}^{+} \\
        \widehat{v}=\frac{1}{s}\psi_{\text{ini}}\text{,} && \text{on }\Omega_{\text{in}}\times\left\{ 0\right\} \\
        is\widehat{w}+\frac{\partial^{2}\widehat{w}}{\partial x^{2}}=\mathcal{L}\left(V\left(\eta\right)w\right)\text{,} && \text{on }\Omega_{\text{out}}\times\mathbb{R}_{0}^{+} \\ 
        \widehat{w}=g\left(\eta\right)\text{,} && \text{on }\left\{ x_{l},x_{r}\right\} \times\mathbb{R}_{0}^{+}, \\ \underset{\left|x\right|\rightarrow0}{\lim}\widehat{w}\left(x,t\right)=0\text{,}\text{for all }t\in\mathbb{R}^{+} \\ 
        \widehat{w}=0\text{,} && \text{on }\Omega_{\text{out}}\times\left\{ 0\right\}. \label{semi-spectral-problem}
    \end{aligned}
\end{equation}
Due to the results of \cite{Ehrhard08, Zheng08} the formulation (\ref{semi-spectral-problem}) can also be derived for periodic potentials as well. For these types of potentials, the semi-spectral variant of the optimization problem (\ref{semi-spectral-problem}) will produce straightforward algebraic expressions in place of these potentials. 

Furthermore, one may reformulate the control and cost functional using the uni-lateral Laplace transform resulting in
\begin{equation}
\underset{\widehat{\eta}}{\min}\frac{\alpha}{p}\underset{\Omega_{\text{in}}}{\int}\left|\widehat{v}\left(s,T;\widehat{\eta}\right)\left(x\right)-\mathcal{L}\left(\overline{\psi}\left(x\right)\right)\right|^{p}dx+\frac{\beta}{q}\underset{\mathbb{R}\times\mathbb{R}_{0}^{+}}{\int}\widehat{\mu}\left(\widehat{\eta}\left(x,t\right)\right)^{q}d\left(x,t\right),
\end{equation}
where $\widehat{\eta}$ denotes the control in frequency space and $\widehat{\mu}$ is an alternative measure to attribute cost to the control.

\section{Examples}
This section is dedicated to working out a couple of examples that show how the presented technique can be used to achieve desired states for the transmon \cite{Roth22, Roth23, Koch07} and fluxonium \cite{Koch09, Bao2021FluxoniumAA} qubits. First, the specification of the system to qubits of the transmon and fluxonium kind is discussed. Dependent on the aforementioned state equations, a method that can be used to translate binary data into this setting is detailed.

\subsection{The Motivation for Considering Optimization in Frequency Space}
The following example illustrates the possible benefits of frequency space optimization. Many quantum algorithms require an even superposition as the initial state on which the computation is performed. For a two-qubit system, this state may be represented as
\begin{equation}
\left|\psi_{\text{initial}}\right\rangle =\frac{1}{2}\left(\left|00\right\rangle +\left|01\right\rangle +\left|10\right\rangle +\left|11\right\rangle \right),
\end{equation}
where $\left|\psi_{\text{initial}}\right\rangle $ denotes a target state for the optimization (which is the initial state for the actual quantum computation). The matrix for the Quantum Fourier Transform with respect to the canonical basis for a two qubit system reads
\begin{equation}
F_{2}=\frac{1}{2}\left[\begin{array}{cccc}
1 & 1 & 1 & 1\\
1 & i & -1 & -i\\
1 & -1 & 1 & -1\\
1 & -i & -1 & i
\end{array}\right].
\end{equation}
One may notice that 
\begin{equation}
F_{2}\left|\psi_{\text{initial}}\right\rangle =\left|00\right\rangle,     
\end{equation}
which reveals that the target vector in frequency space may be captured with the use of only a single basis vector instead of four. Assuming now that we include only this basis vector from the frequency space, which already contains the target state, we may ``add harmonics", i.e. vectors that form an orthogonal basis which include 
\begin{equation}
\left|\psi_{\text{initial}}\right\rangle =\left(\frac{1}{2}\left[\begin{array}{c}
1\\
1\\
1\\
1
\end{array}\right]\right)_{\text{canonical}}=\left(\left[\begin{array}{c}
1\\
0\\
0\\
0
\end{array}\right]\right)_{\text{frequency space}}.
\end{equation}
Using this process, we may control the numerical extent of the problem, while always including the target state. In many cases, quantum systems exhibit symmetries within the frequency domain, and this results in a small number of harmonics being sufficient to solve the problem accurately.

\subsection{The Transmon Qubit}
A common formulation of the Hamiltonian for a superconducting transmon qubit reads
\begin{equation}
H=4E_{C}\left(\widehat{n}-n_{g}\right)^{2}-E_{J}\cos\left(\widehat{\varphi}\right),    
\end{equation}
where the observables are the phase operator $\widehat{\varphi}$ and the charge operator $\widehat{n}$ satisfying the canonical commutation relation $\left[\widehat{n},\widehat{\varphi}\right]=\hbar i$. The value $n_{g}$ gives the equilibrium charge value, the charging and Josephson energies are denoted by $E_{C}$ and $E_{J}$ respectively while the electron charge $e$ and total capacitance $C_{\varSigma}$ relate with the charging energy by
\begin{equation}
    E_{C}=\frac{e^{2}}{2C_{\varSigma}}.
\end{equation}
Relating this formulation to the preceding chapters, one can also split the Hamiltonian into the potential and momentum operator parts by defining
\begin{equation}
    P\left(\widehat{n}\right)=4E_{C}\left(\widehat{n}-n_{g}\right)^{2},\quad Q\left(\widehat{\varphi}\right)=-E_{J}\cos\left(\widehat{\varphi}\right),
\end{equation}
which allows us to write
\begin{equation}
    H=P\left(\widehat{n}\right)+Q\left(\widehat{\varphi}\right).
\end{equation}
To finish the formulation of the Schrödinger equation in coordinate space, a transformation into a ``phase basis" is considered. Using this technique, the phase operator $\widehat{\varphi}$ becomes a position variable and the charge operator $\widehat{n}$ becomes $-i\frac{\partial}{\partial\varphi}$. This effort culminates in a form of the Schrödinger equation, which reads
\begin{equation}
    \left[4E_{C}\left(-i\frac{\partial}{\partial\varphi}-n_{g}\right)^{2}-E_{J}\cos\left(\varphi\right)\right]\psi\left(\varphi,t\right)=i\hbar\frac{\partial}{\partial t}\psi\left(\varphi,t\right),\label{schrodinger_transmon}
\end{equation}
where the (complex-valued) wave function additionally satisfies a periodic condition \cite{Koch07} $\psi\left(\varphi,t\right)=\psi\left(\varphi+2k\pi,t\right)$.
According to \cite{Roth23} it is possible to engineer the surrounding circuitry of a transmon qubit in such a way that $E_{C}$ and $E_{J}$ may be controlled within a given range. The formulation of the transparent boundary conditions for (\ref{schrodinger_transmon}) can be made as \cite{roth2023maxwellschrodinger}. This directly leads to a control problem of type (\ref{the_main_optimization_problem}) or (\ref{semi-spectral-problem}), where the control set is $\theta=E_{J}$. As mentioned in Section \ref{sec_extensions}, we can extend the formulation to include the control of $E_{C}$, resulting in the control set $\theta=\left(E_{J},E_{C}\right)$, which leads to the ability to control the ratio $\frac{E_{J}}{E_{C}}$, which impacts the energy spectrum, especially at higher levels.

\subsection{The Fluxonium Qubit}
Following the same ideas, the fluxonium qubit \cite{bao2022fluxonium} can be treated. In contrast to its transmon counterpart, the fluxonium qubit has an additional control parameter, induced by a magnetic field, which is the consequence of adding multiple Josephson junctions into the circuit \cite{Koch09, Bao2021FluxoniumAA}. The Hamiltonian in this case reads
\begin{equation}
H=4E_{C}\left(\widehat{n}-n_{g}\right)^{2}-E_{J}\cos\left(\widehat{\varphi}+\varphi_{\text{ext}}\right)+\frac{1}{2}E_{L}\widehat{\varphi}^{2},    
\end{equation}
where $\varphi_{\text{ext}}$ relates to the external magnetic flux and $E_{L}$ is the inductive energy of the qubit. Just like in the case of the transmon qubit, the Schrödinger equation can be derived and reads
\begin{equation}
\left[4E_{C}\left(-i\frac{\partial}{\partial\varphi}-n_{g}\right)^{2}-E_{J}\cos\left(\varphi+\varphi_{\text{ext}}\right)+\frac{1}{2}E_{L}\varphi^{2}\right]\psi\left(\varphi,t\right)=i\hbar\frac{\partial}{\partial t}\psi\left(\varphi,t\right).    
\end{equation}
Treating the boundary conditions as before, we arrive at a control problem formulation with control sets $\theta=\left(E_{J}\right)$ or $\theta=\left(E_{J},E_{C},\varphi_{\text{ext}}\right)$ depending on the whether the approach of Section \ref{sec_extensions} is applied, or not.

\subsection{A Possible Encoding of Data - Preparing an Initial State for Quantum Computation}
To detail the application of the proposed frequency domain-based optimization a particular example of achieving an initial state for a computation on a quantum computer is presented. This example serves only as an illustration of a possible application and does not exclude the application of the method to different quantum systems, in which the initial data may not be subject to the qubit interpretation.

For clarity, the procedure is detailed using a single qubit system, but extends readily to a multi qubit system, which is discussed subsequently. Let $\left|\psi_{\text{target}}\right\rangle \in\mathbb{C}^{2}$ be the target state satisfying the usual norm condition $\left\Vert \left|\psi_{\text{target}}\right\rangle \right\Vert _{\mathbb{C}^{2}}=1$. More explicitly, one may then assume that there exist $\alpha,\beta\in\mathbb{C}$ such that 
\begin{equation}\label{eq_schrodinger_transmon}
\left|\psi_{\text{target}}\right\rangle =\alpha\left|0\right\rangle +\beta\left|1\right\rangle ,
\end{equation}
where the norm condition reads $\left|\alpha\right|^{2}+\left|\beta\right|^{2}=1$ and $\left|0\right\rangle , \left|1\right\rangle $ are a basis of $\mathbb{C}^{2}$.

It is known \cite{bardin20} that transmon qubit systems decribed by \eqref{eq_schrodinger_transmon} has a discrete energy spectrum and its eigenvalues may be approximated
\begin{equation}\label{eq_spectrum_of_trans}
E_{n}\approx\hbar\omega_{0}\left(n+\frac{1}{2}\right)-\frac{E_{c}}{12}\left(6n^{2}+6n+3\right),
\end{equation}
where $\hbar$ denotes the Planck constant. Due to the form of \eqref{eq_spectrum_of_trans} it is possible to manipulate the spacing between the energy levels by means of $E_{c}$. Any state $\left|\varphi\right\rangle$ may be written in the countable eigenbasis as
\begin{equation}\label{eq_inf_dim_state}
\left|\varphi\right\rangle =\sum_{i=0}^{\infty}a_{i}\left|i\right\rangle,
\end{equation}
where $\left|i\right\rangle $ are the eigenstates of the Hamiltonian. It is common practice to construct transmon qubits, such that only the first two energy levels are accessible, which validates the use of the two dimensional model, which makes use of the space $\mathbb{C}^{2}$ instead of the infinite dimensional separable Hilbert space in which the state \eqref{eq_inf_dim_state} is found. 

Using these assumptions and the fact that the state \eqref{eq_schrodinger_transmon} may also be considered to be an element of this larger Hilbert space if we interpret the basis vectors in \eqref{eq_schrodinger_transmon} as eigenvectors of the Hamiltonian, the target state represents a valid element within the solution space of the Schrodinger equation. Applying the Laplace transform
\begin{equation}
 \overline{\psi}=\mathcal{L}^{-1}\left(\psi_{\text{initial}}\right)=\alpha\mathcal{L}^{-1}\left(\left|0\right\rangle \right)+\beta\mathcal{L}^{-1}\left(\left|1\right\rangle \right)   
\end{equation}

yields the target state for the semi-spectral problem \eqref{semi-spectral-problem}.

To extend this construction to the case of the multi qubit setting, one only needs to consider target states that are a tensor product of constituent states and consider a tensor product solution space for the Schrödinger equation also. In this setting, the aforementioned steps may be repeated.

\section{Conclusion}
A semi-spectral method for optimizing state preparation was presented. Due to the chosen method of derivation, the artificial boundary conditions were shown to be directly translatable to the semi-spectral form. This method directly generalizes to other types of potentials outside of the domain, especially periodic ones, which due to the spectral formulation of the optimization problem (\ref{semi-spectral-problem}) will have a simple form.

\hskip 2cm
\paragraph*{Acknowledgement}
Ales Wodecki acknowledges the support of the Czech
Science Foundation (23-07947S). \\

\paragraph*{Disclaimer} 
This paper was prepared for information purposes and
is not a product of HSBC Bank Plc. or its affiliates.
Neither HSBC Bank Plc. nor any of its affiliates make
any explicit or implied representation or warranty, and
none of them accept any liability in connection with this
paper, including, but not limited to, the completeness,
accuracy, reliability of information contained herein and
the potential legal, compliance, tax, or accounting effects
thereof. This document is not intended as investment
research or investment advice, or a recommendation, offer
or solicitation for the purchase or sale of any security,
financial instrument, financial product or service, or to be
used in any way for evaluating the merits of participating
in any transaction.

\clearpage
\bibliographystyle{siamplain}
\bibliography{references}

\end{document}